\documentclass[twocolumn,aps,prl,groupedaddress,nofootinbib,showpacs]{revtex4}
\usepackage{graphicx}
\DeclareGraphicsExtensions{.eps}
\usepackage{amsmath}
\usepackage{bm}
\usepackage{slashed}
\usepackage{epsfig}
\usepackage{amsfonts}

\newcommand\mev{\mathrm{~MeV}}
\newcommand\gev{\mathrm{~GeV}}
\newcommand\tev{\mathrm{~TeV}}

\newcommand{\jpsi}{{J/\psi}}

\newcommand{\state}[4]{{^#1\hspace{-0.6mm}#2_{#3}^{[#4]}}}

\newcommand\CScSa{\state{3}{S}{1}{1}}

\newcommand\COaSz{\state{1}{S}{0}{8}}

\newcommand\COcSa{\state{3}{S}{1}{8}}

\newcommand\COcPj{\state{3}{P}{J}{8}}

\newcommand{\ben}{\begin{eqnarray}}
\newcommand{\een}{\end{eqnarray}}
\newcommand{\nnu}{\nonumber\\}
\newcommand{\bef}{\begin{figure}[!htp]}
\newcommand{\eef}{\end{figure}}

\newcommand{\bea}{\begin{eqnarray}}
\newcommand{\eea}{\end{eqnarray}}

\def\ba{\begin{linenomath*}\begin{equation}}
\def\ea{\end{equation}\end{linenomath*}}

\begin{document}
\title{Factorized power expansion for high-$p_T$ heavy quarkonium production}

\author{Yan-Qing Ma$^1$,
Jian-Wei Qiu$^{1,2,3}$, George Sterman$^{2,3}$ and Hong Zhang$^{3}$}

\email{yqma@bnl.gov, jqiu@bnl.gov, sterman@insti.physics.sunysb.edu, hong.zhang@stonybrook.edu}

\affiliation{$^1$Physics Department,
                Brookhaven National Laboratory,
                Upton, NY 11973-5000, USA}

\affiliation{$^2$C.N.\ Yang Institute for Theoretical Physics,
             Stony Brook University,
             Stony Brook, NY 11794-3840, USA}

\affiliation{$^3$Department of Physics and Astronomy, Stony Brook University, Stony Brook, NY 11794, USA}

\date{\today}

\begin{abstract}
We show that when the factorized cross section for heavy quarkonium production includes next-to-leading power (NLP) contributions associated with the production of the heavy quark pair at short distances, it naturally reproduces all high $p_T$ results calculated in nonrelativistic QCD (NRQCD) factorization.    This extended formalism requires fragmentation functions for heavy quark pairs, as well as for light partons.  When these fragmentation functions are themselves calculated using NRQCD, we find that two of the four leading NRQCD production channels, $\CScSa$ and $\COaSz$, are dominated by the NLP contributions for a very wide $p_T$ range.
The large next-to-leading order corrections of NRQCD are absorbed into the leading order of the first power correction.     The impact of this finding on heavy quarkonium production and its polarization is discussed.
\end{abstract}

\maketitle


\allowdisplaybreaks

{\it Introduction.}---  The conjecture of nonrelativistic QCD (NRQCD) factorization for heavy quarkonium production~\cite{Bodwin:1994jh} has proved  quite successful phenomenologically~\cite{Brambilla:2010cs}, once next-to-leading order (NLO) corrections are included.   The relatively recent NRQCD calculation for $\jpsi$ production at hadron colliders at NLO, which took several groups several years to complete~\cite{Butenschoen:2012px,Chao:2012iv,Gong:2012ug}, does a much better job in fitting the data than does the leading order (LO) approximation, although it still relies on only four leading channels: $\CScSa$, $\COaSz$, $\COcSa$ and $\COcPj$  (in standard spectroscopic notation for the angular momentum states of the produced heavy quark pair, and with superscripts $1$ and $8$ for the color-singlet and color-octet states, respectively).  In particular, the NLO calculation provides a potential solution for the long-standing polarization puzzle at high $p_T$~\cite{Chao:2012iv,Bodwin:2014gia,Faccioli:2014cqa}.  At the same time, the NLO NRQCD calculation does not yet give a fully satisfactory picture of heavy quarkonium production.  Some channels give NLO corrections that are orders larger than the corresponding LO results.    This raises concerns on the stability of the expansion in $\alpha_s$, while it would be extremely difficult  to go beyond NLO.  In addition, the lack of analytic expressions at NLO makes it difficult to obtain a clear physical picture on how the various NRQCD channels of the heavy quark pair are actually produced.

Recently, a systematic QCD factorization approach to high $p_T$ heavy quarkonium production was described \cite{Kang:2014tta,KMQS-hq2}, based on related earlier work~\cite{Nayak:2005rw,Nayak:2005rt,Kang:2011mg,Kang:2011zza}. A similar factorization formalism has also been derived from soft-collinear effective theory~\cite{Fleming:2012wy,Fleming:2013qu}. In this approach, the production cross section is expanded in powers of $1/p_T^2$ first, and then in powers of $\alpha_s$, in contrast to conventional NRQCD factorization, which includes all power contributions at each power of $\alpha_s$.  Large logarithmic contributions can be resummed systematically  by solving a closed set of evolution equations \cite{Kang:2014tta}.  The leading power (LP) terms are given by the production of a single active parton at a distance scale of ${\cal O}(1/p_T)$, which then fragments into an observed heavy quarkonium \cite{Nayak:2005rw}.  The next-to-leading power (NLP) contribution is dominated by the perturbative  production of a heavy quark-antiquark pair at short distances, followed by fragmentation of the pair into a heavy quarkonium.   This requires a new set of quark-pair fragmentation functions (FFs), which have been defined in Ref.\ \cite{Kang:2014tta}.

In this Letter, we compare the calculation of $\jpsi$ production for the factorizated power expansion, including the leading and the first nonleading powers in $p_T$, to the NLO NRQCD calculation.  We find that the large NLO contributions to the $\CScSa$ and $\COaSz$ channels in NRQCD
are due primarily to LO NLP corrections in the factorized power expansion.    Specifically, with heavy quark pair FFs calculated in NRQCD \cite{Ma:2013yla,Ma:2014eja}, the LO contribution at NLP nicely reproduces NLO NRQCD results for both $\CScSa$ and $\COaSz$ channels for a wide $p_T$ range.      These results are independent of specific fits to matrix elements.   At the same time, several recent global fits to Tevatron and LHC quarkonium production data agree on an important, and in some cases dominant, role for these channels, compared to  LP contributions from $\COcSa$ and $\COcPj$, over a wide range of $p_T$ \cite{Ma:2010yw,Ma:2010jj,Bodwin:2014gia,Faccioli:2014cqa}.   Thus, the results of this paper show that a systematic treatment of NLP contributions is mandatory for understanding heavy quarkonium production at today's collider energies.  In addition, with the factorization of NLP contributions, we gain control in principle over the class of higher order corrections that describe the evolution of the heavy quark pair and its mixing with light partons.

{\it QCD factorized power expansion.}---The production of a heavy quarkonium $H$ is factorized as \cite{Kang:2014tta}
\begin{align}\label{eq:factorization}
&\hspace{-0.2in}d\sigma_{A+B\to H+X}(p)\nnu
\approx&\sum_f \int_0^1 \frac{dz}{z^2} D_{f\to H}(z)\, d\hat{\sigma}_{A+B\to f(p_c)+X}(p/z)
\nnu
&+\sum_\kappa \int_0^1 \frac{dz}{z^2}\int_{-1}^{1} \frac{d\zeta_1 d\zeta_2}{4}\,
{\cal D}_{[Q\bar{Q}(\kappa)]\to H}(z,\zeta_1,\zeta_2)
\\
&\times d\hat{\sigma}_{A+B\to [Q\bar{Q}(\kappa)](p_c)+X}(p(1\pm\zeta_1)/2z,p(1\pm\zeta_2)/2z),
\nonumber
\end{align}
where $D_{f\to H}$ (${\cal D}_{[Q\bar{Q}(\kappa)]\to H}$) are single (double) parton FFs, which give the LP (NLP) contribution, $\sum_f$ runs over all parton flavor $f=q,\bar{q},g$, and $\sum_\kappa$ includes all spin and color states of fragmenting heavy quark-antiquark pairs: $v^{[1,8]}$, $a^{[1,8]}$, or $t^{[1,8]}$, where $v$, $a$ and $t$ refer to the vector, axial-vector, and tensor states of the pair's spin.  In Eq.~(\ref{eq:factorization}), the $d\hat\sigma$ are short-distance coefficients (SDCs) to produce on-shell fragmenting parton(s), and contain all information about the initial colliding state, including convolutions with parton distribution functions (PDFs) if $A$ and $B$ are hadrons. Longitudinal momentum fractions are defined as $z=p^+/p_c^+$, $\zeta_1=2q_1^+/p_c^+$ and $\zeta_2=2q_2^{+}/p_c^+$, where $p^+$, $p_c^+$, and $q_1^+$ ($q_2^{+}$) are the light-cone ``$+$'' components of, respectively, the momenta of the quarkonium, the fragmenting single parton or heavy quark-antiquark pair, and half the relative momentum of the heavy quark and antiquark in the amplitude (complex conjugate amplitude).

The predictive power of the factorization formalism in Eq.~(\ref{eq:factorization}) relies on the SDCs and our knowledge of FFs.  The SDCs for producing a single parton are known to NLO \cite{Aversa:1988vb}, and the complete LO SDCs for producing a heavy quark-antiquark pair have also been calculated \cite{Kang:2011mg,KMQS-hq2}.  Although the LO evolution kernels of FFs are available \cite{Kang:2014tta}, we still need input FFs at a scale $\mu_0\gtrsim 2m_Q$ with heavy quark mass $m_Q\gg \Lambda_{\rm QCD}$.  The input FFs are nonperturbative and, in principle, must be extracted from data. Extracting several three-variable input FFs, however, may not be an easy task in practice.
Nevertheless, when its invariant mass is sufficiently near the input scale,
the heavy quark-antiquark pair is effectively a nonrelativistic system in its rest frame.  Then, applying NRQCD factorization as a natural conjecture at this scale, all relevant input FFs to a heavy quarkonium can be evaluated analytically, and expressed in terms of a few universal NRQCD long-distance matrix elements (LDMEs) with perturbative coefficients \cite{Kang:2011mg},
\begin{align}\label{eq:nrqcdffs}
&\hspace{-0.2in}
{\cal D}_{[Q\bar{Q}(\kappa)]\to H}(z,\zeta_1,\zeta_2;m_Q,\mu_0)
\nnu
=&
\sum_{[Q\bar{Q}(n)]}\hat{d}_{[Q\bar{Q}(\kappa)]\to [Q\bar{Q}(n)]}(z,\zeta_1,\zeta_2;m_Q,\mu_0,\mu_\Lambda) \nnu
& \hspace{0.4in}
\times \langle {\cal O}_{[Q\bar{Q}(n)]}^H(\mu_\Lambda)\rangle\, ,
\end{align}
where $[Q\bar{Q}(n)]$ are NRQCD heavy quark-antiquark states, $\mu_\Lambda \sim {\cal O}(m_Q)$ is the NRQCD factorization scale, and $\langle {\cal O}_{[Q\bar{Q}(n)]}^H(\mu_\Lambda)\rangle$ are the LDMEs.  The perturbative coefficients of the FFs, the $\hat{d}$'s in Eq.~(\ref{eq:nrqcdffs}), are available to LO and NLO in $\alpha_s$ \cite{Ma:2013yla,Ma:2014eja}.  Although a formal proof of NRQCD factorization is still lacking, these calculated FFs should be a reasonable approximation to the  true input FFs, in view of the phenomenological successes of NRQCD factorization. With the FFs calculated in NRQCD, the factorized power expansion in Eq.~\eqref{eq:factorization} and conventional NRQCD factorization are mutually consistent at LP and NLP when summed to all orders in $\alpha_s$.

For numerical predictions, we need to use the SDCs and evolution kernels of FFs and PDFs at the same order in their perturbative expansion to provide a fully consistent factorized cross section.  For example, for the LO contribution to the cross section, we should use LO PDFs and FFs (evaluated with LO kernels) and LO SDCs.  It is important to note, however, that the order of the  factorized cross section should be distinguished from the order at which we calculate the input FFs in NRQCD.  To have the best model predictions for the FFs, we should always use the input FFs calculated in NRQCD at the highest order available in $\alpha_s$ evaluated at the NRQCD factorization scale, regardless the order at which we evaluate the perturbative contribution to the factorized cross section.

{\it Comparison with NLO NRQCD cross sections.}---  We show in this subsection that the bulk of the terribly complicated NLO NRQCD results for high-$p_T$ hadronic $\jpsi$ production \cite{Butenschoen:2012px,Chao:2012iv,Gong:2012ug} are reproduced by analytic LO contributions in the factorized power expansion. More precisely, we compare the NLO NRQCD results of Refs.~\cite{Ma:2010yw,Ma:2010jj} with predictions of Eq.~\eqref{eq:factorization}, using the FFs calculated in NRQCD, for all four leading, $\CScSa$, $\COaSz$, $\COcSa$ and $\COcPj$, NRQCD production channels.

The factorized power expansion in Eq.~(\ref{eq:factorization}) and NRQCD factorization organize the order of perturbative contributions to heavy quarkonium production differently.  In particular, in the power expansion  we must independently specify the order of evolution for parton distributions and the order at which we compute FFs as well as SDCs.  Our choices for this numerical comparison are listed in Table~\ref{tab:parameter}.      To compare our LO predictions with NLO NRQCD calculations, we in general evaluate Eq.~(\ref{eq:factorization}) for both the LP and NLP contributions with the LO hard parts \cite{KMQS-hq2}, LO PDFs (CTEQ6L1 \cite{Pumplin:2002vw}), and FFs  from Refs.~\cite{Ma:2013yla,Ma:2014eja} without including the evolution.   We use, however, NLO PDFs for the $\COaSz$ and  $\COcPj$ channels at LP because the NLO NRQCD calculation for these channels  uses NLO PDFs. The FFs  are calculated using the method in \cite{Ma:2013yla,Ma:2014eja} so that their order added to the order of the SDCs matches the order of the corresponding NLO NRQCD calculations.  A summary of available single parton to heavy quarkonium FFs, which are needed for the LP contribution, can be found in Ref.~\cite{Ma:2013yla} and references therein.
\begin{table}[h]
\caption{\label{tab:parameter} The choices for our LO QCD factorization calculations in Eq.~(\ref{eq:factorization}) for comparison with NLO NRQCD calculations.   Explicit formulas for the FFs can be found in Ref.\ \cite{Ma:2013yla,Ma:2014eja}, the SDCs in Ref.\ \cite{Kang:2011mg,KMQS-hq2} for NLP and Ref.\ \cite{Aversa:1988vb} for LP.}
\begin{tabular}{c|cc|cc|cc|cc}
\hline\hline \itshape ~\rm{Channel}~ & $\CScSa$&$\CScSa$& $\COcSa$ &$\COcSa$& $\COaSz$&$\COaSz$& $\COcPj$&$\COcPj$
\\ ~Power~ & ~LP~ & NLP& ~LP~ & NLP& ~LP~ & NLP& ~LP~ & NLP
\\\hline ~PDFs~&~-~& LO & LO & LO~&~NLO~& LO & NLO & LO
\\ ~FFs~&~-~& $\alpha_s^1$ & $\alpha_s^1$ & $\alpha_s^0$~&~$\alpha_s^2$~& $\alpha_s^0$ & $\alpha_s^2$ & $\alpha_s^0$
\\~SDCs~&~-~& $\alpha_s^3$ & $\alpha_s^2$ & $\alpha_s^3$~&~$\alpha_s^2$~& $\alpha_s^3$ & $\alpha_s^2$ & $\alpha_s^3$
\\\hline\hline
\end{tabular}
\end{table}

\begin{figure}[!tbhp]
\begin{center}
\includegraphics*[scale=0.85]{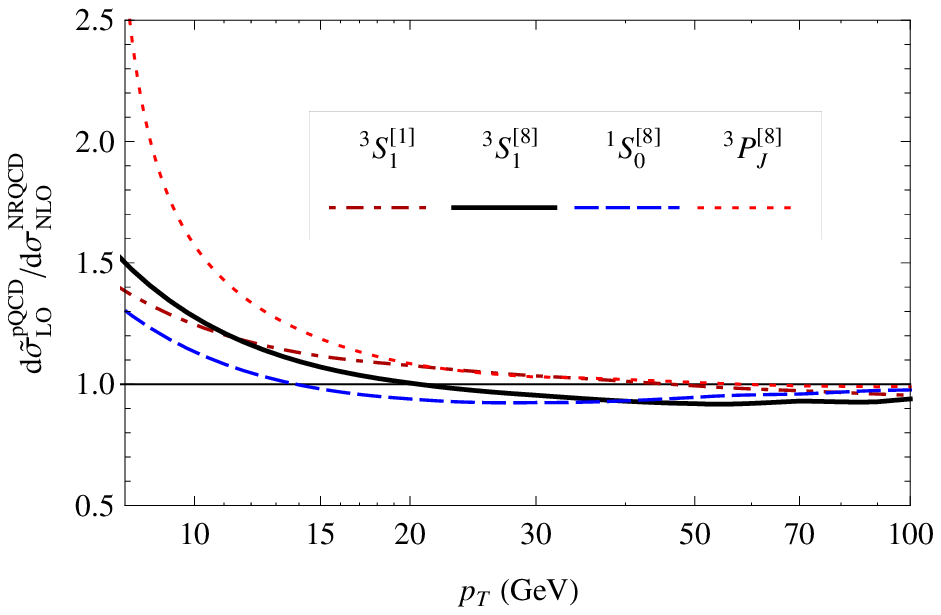}
\caption{Ratio of $\jpsi$ production rate from LO QCD factorization over that of NLO NRQCD calculation for four leading NRQCD channels. See the text for details.}
 \label{fig:ratioNLONRQCD}
 \end{center}
\end{figure}

In Fig.~\ref{fig:ratioNLONRQCD}, we show the ratios of our analytic LO predictions with the PDFs and parameter choices in Table~\ref{tab:parameter} to the numerical results of NLO NRQCD calculations in various leading NRQCD channels.  Note that the values of the NRQCD matrix elements cancel in these ratios.    The tilde of $d\tilde{\sigma}^{\rm QCD}_{\rm LO}$ indicates the slightly modified LO contribution, with the choices specified in Table~\ref{tab:parameter}, to better match the NLO NRQCD calculations.  We choose $\sqrt{S}=7\tev$ and $|y|<0.9$ for a typical kinematic regime at the LHC.  We take charm quark mass $m_c=1.5 \gev$, $\Lambda_{\text{QCD}}^{(5)}=165\mev$ ($\Lambda_{\text{QCD}}^{(5)}=226\mev$) for LO (NLO) $\alpha_s$ with quark active flavors $n_f=5$, and CTEQ6M when NLO PDFs are needed \cite{Pumplin:2002vw}, and set the renormalization, factorization, and the NRQCD scales to $\mu_r=\mu_f=p_T$ and $\mu_\Lambda=m_c$, respectively.

As shown in Fig.~\ref{fig:ratioNLONRQCD}, our slightly modified LO QCD calculation can almost reproduce the NLO NRQCD calculation channel by channel for $p_T>10-15\gev$, depending on the channel.  The comparison in Fig.~\ref{fig:ratioNLONRQCD} demonstrates that the very complicated results of NLO NRQCD calculations can be reproduced by the simple and fully analytic LO calculation of the QCD factorization approach for $p_T>10\gev$, and clearly indicates that perturbative organization of the factorized power expansion is well suited to heavy quarkonium production at high $p_T$.  It also shows the importance of the NLP contribution.  Without it, as shown in Ref.~\cite{Bodwin:2014gia}, for example, the LP QCD factorization contribution can only reproduce NLO NRQCD results for the $\COcSa$ and $\COcPj$ channels at large $p_T$, not for the $\CScSa$ and $\COaSz$ channels.
\begin{figure}[!tbhp]
\begin{center}
\includegraphics*[scale=0.85]{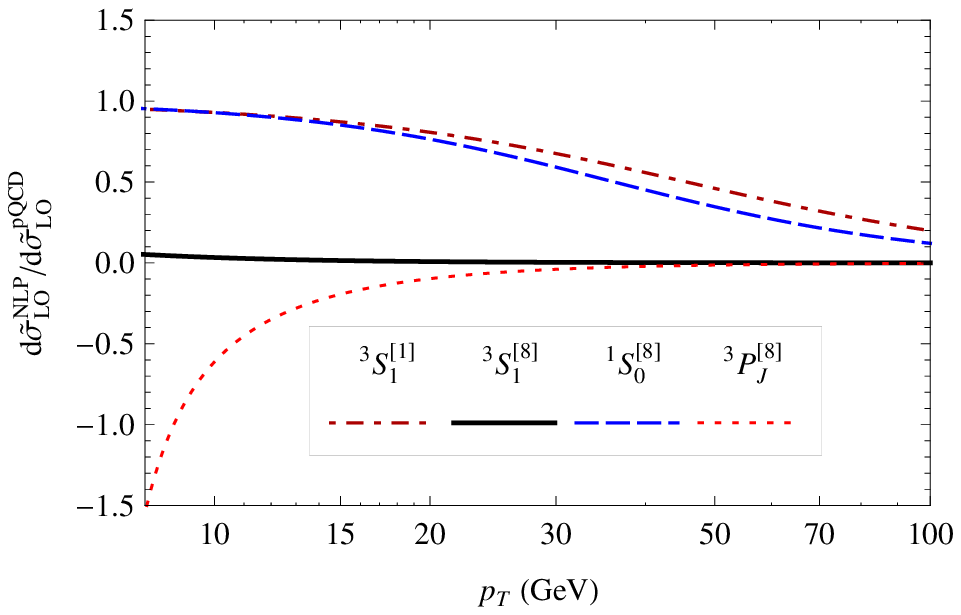}
\caption{Ratio of NLP contributions to total contribution in LO QCD for each channel. $d\tilde{\sigma}$ means we have a special choice for PDFs. See text for details.}
 \label{fig:ratioNLP}
 \end{center}
\end{figure}

To further illustrate the importance of NLP contributions, we plot the ratio of the NLP contribution to the total LO QCD contribution in Fig.~\ref{fig:ratioNLP} for each channel.  Figure~\ref{fig:ratioNLP} clearly shows that NLP contributions are negligible for the $\COcSa$ channel over the full $ p_T$ range, and are small for the $\COcPj$ channel when $p_T>20\gev$, beyond which it is below $10$ percent. However, the NLP contributions are crucial for $\COaSz$ and $\CScSa$ channels even if $p_T$ approaches 100 $\gev$.  Since the FFs for a single active parton to fragment into a $\CScSa$ heavy quark pair, calculated in NRQCD, vanish for both LO and NLO, as indicated in Table \ref{tab:parameter} the two-loop gluon FF derived in Refs.~\cite{Braaten:1993rw,Bodwin:2012xc} was used for the LP contribution to the $\CScSa$ channel in Fig.~\ref{fig:ratioNLP}.

In the above comparison with NLO NRQCD calculations, we did not include the evolution of FFs.  A complete LO QCD calculation should include the evolution of FFs using the LO evolution kernels given in Ref.\ \cite{Kang:2014tta} and input FFs calculated in NRQCD factorization at NLO \cite{Ma:2013yla,Ma:2014eja}, and a set of updated NRQCD LDMEs by fitting the data.  From its consistency with the existing NLO NRQCD results, and the control through evolution of its higher order corrections, we expect such a  LO QCD factorized power expansion to clarify existing data on heavy quarkonium production at collider energies.  Also, because the LP  $\COcSa$ and $\COcPj$ channels, which produce predominantly transversely polarized heavy quarkonia, appear not to be dominant \cite{Chao:2012iv,Bodwin:2014gia,Faccioli:2014cqa}, heavy quarkonium production at current collider energies is strongly influenced by the $\COaSz$ channel, and is more likely to be unpolarized.

{\it Heavy quark-antiquark FFs.}--- To evaluate the factorization formula in Eq.~(\ref{eq:factorization}), we have employed FFs calculated in NRQCD factorization \cite{Ma:2013yla,Ma:2014eja}.  The perturbative coefficients of these calculations, $\hat{d}$'s, Eq.~(\ref{eq:nrqcdffs}), are distributions, defined under the integration and expressed in terms of $\delta$ functions and the $+$ prescriptions.  Consequently, unlike the FFs of light hadrons extracted from the data,  FFs calculated in NRQCD are not smooth functions of momentum fractions.

On the other hand, both the SDCs in Eq.~(\ref{eq:factorization}) and the perturbative coefficients of (\ref{eq:nrqcdffs}) are known analytically.  It is very easy to track down the dependence on each of the convolution variables---the momentum fractions $z$, $\zeta_1$ and $\zeta_2$.  In particular, the LO  contribution to each input FF calculated in NRQCD is proportional to the product $\delta(\zeta_1)\delta(\zeta_2)$ and its derivatives.  Higher order corrections from QCD evolution allow the exchange of momentum between the active heavy quark and antiquark, so that at order $\alpha_s$ the range of $|\zeta_1|$ and $|\zeta_2|$ is proportional to the phase space available for gluon radiation.  That is, the effective range of $|\zeta_1|$ and $|\zeta_2|$ is limited by $1-z$.   Since in general factorized hadronic cross sections are dominated by the large $z$ region of FFs due to the steeply falling PDFs of colliding hadrons \cite{Berger:2001wr}, we expect that the convolution over $\zeta$'s in Eq.~(\ref{eq:factorization}) is  dominated by the region where $\zeta_1\sim \zeta_2\sim 0$.

\begin{figure}[!tbhp]
\begin{center}
\includegraphics*[scale=0.8]{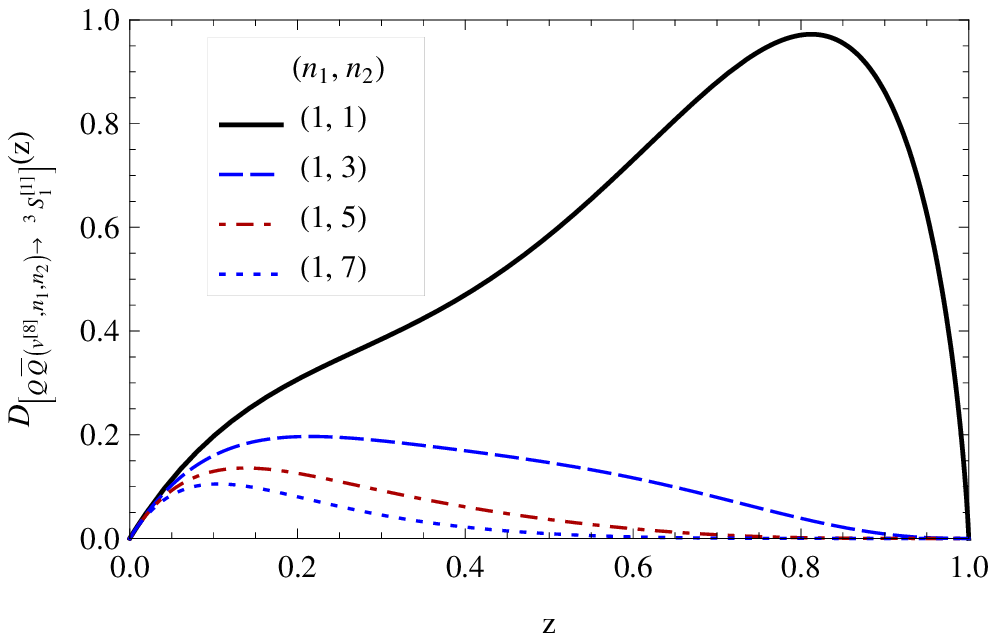}
\caption{First few moments of ${\cal D}^{[n_1,n_2]}_{[c\bar{c}(v^{[8]})]\to \CScSa}(z)$ with an arbitrary normalization.}
 \label{fig:V33S11}
 \end{center}
\end{figure}

To demonstrate this feature explicitly, we take the convolution over $\zeta_1$ as an example (the integration over $\zeta_2$ is equivalent).  At the LO, the most singular $\zeta_1$ dependence of the short-distance  coefficients, $\hat{\sigma}$ in Eq.~(\ref{eq:factorization}), is proportional to $1/(1-\zeta_1^2)$.  The apparent singularity near $\zeta_1=\pm 1$ is the well-known endpoint singularity;  as discussed in the Appendix~A of Ref.~\cite{Kang:2014tta}, it does not cause any divergence to the cross section and can be absorbed into the definition of the fragmentation functions.  Using the FFs calculated in NRQCD \cite{Ma:2013yla,Ma:2014eja}, we have also verified that $\lim_{\zeta_1\to \pm 1}{\cal D}_{[Q\bar{Q}(\kappa)]\to H}(z,\zeta_1,\zeta_2)/(1-\zeta_1^2)$ is finite.  Since $1/(1-\zeta_1^2)=\sum_{n=0}^{\infty}\zeta_1^{2n}$ for $\zeta<1$, and the FFs have a limited range in $\zeta_1$, the convolution over $\zeta_1$ between the hard coefficients and the FFs can be expressed as a sum of the $\zeta_1$ moments of the FFs,
\begin{align}
& {\cal D}^{[n_1,n_2]}(z)
\equiv
\int_{-1}^1\frac{d\zeta_1\, d\zeta_2}{4}
{\zeta_1^{n_1} \zeta_2^{n_2}}{\cal D}(z,\zeta_1,\zeta_2)\, ,
\end{align}
which should give a quantitative measure of the range of $\zeta$ dependence in the FFs. For example, in the calculations following Table \ref{tab:parameter}, the FF ${\cal D}_{[c\bar{c}(v^{[8]})]\to \CScSa}(z,\zeta_1,\zeta_2)$ is computed at order $\alpha_s$, and has support away from $\zeta_1=\zeta_2=0$.     In Fig.~\ref{fig:V33S11}, we plot the first few moments of  ${\cal D}_{[c\bar{c}(v^{[8]})]\to \CScSa}(z,\zeta_1,\zeta_2)$, as a function of $z$.  The FF itself is  an odd function of $\zeta_1$ and $\zeta_2$ \cite{Ma:2013yla}. Consequently, as shown in Fig.~\ref{fig:V33S11}, only moments with odd integers $n_1$ and $n_2$ survive.  For this calculation, we set the input QCD factorization scale as $\mu_0=2m_c$.

From Fig.~\ref{fig:V33S11} it is clear that FFs with higher moments decrease very quickly for moderate and large $z$, due to the expected suppression in powers of $1-z$, while their values are almost $n_i$ independent at small $z$, where $1-z$ is order unity.  All other channels of FFs calculated by assuming NRQCD factorization demonstrate the same feature. That is, numerically, the convolution over $\zeta_1$ and $\zeta_2$ can be approximated by the first few moments of the fragmentation functions.  We note that like other QCD factorizations, the perturbatively calculated coefficient functions and evolution kernels include ``threshold"  powers of $\log(1-z)$.  In addition, there are also threshold logarithms in powers of $\log(1-\zeta_1^2)$ and $\log(1-\zeta_2^2)$ for the NLP contributions, which are the subject
of ongoing study.

{\it Summary.} We have shown that the LO contribution to hadronic $\jpsi$ production, calculated in a factorized expansion at LP and NLP, naturally reproduces all NLO results calculated in NRQCD factorization for $p_T \gtrsim 10 \gev$.  With the FFs calculated assuming  NRQCD factorization at an input scale of the order of $m_Q$,   NLP contributions are important, and potentially dominant in the production of heavy quarkonia at the current collider energies, at least for the $\CScSa$ and $\COaSz$ channels.  The NLP contribution to the $\COaSz$ channel may dominate the total production rate if, as indicated by recent studies \cite{Bodwin:2014gia,Ma:2010yw,Ma:2010jj},  the sum of LP contributions from $\COcSa$ and $\COcPj$ is relatively small.  If this is indeed the case, the asymptotic transverse polarization of $\jpsi$  \cite{Brambilla:2010cs} will require even higher $p_T$ to set in, and the theory will naturally accommodate  unpolarized or slightly longitudinally polarized cross sections over a wide range of $p_T$.   A more detailed global study and refitting of NRQCD matrix elements for $\jpsi$ cross sections and polarization using the factorized power expansion is clearly necessary, and will require the application of the moment method that we have sketched above.  After almost forty years since the discovery of the $\jpsi$ \cite{Aubert:1974js,Augustin:1974xw}, the production of heavy quarkonia remains one of the most active and fascinating subjects in strong interaction physics, and major progress has been made in last decade \cite{Brambilla:2010cs,Bodwin:2013nua}.


We thank G.T.~Bodwin and Z.-B.~Kang for many helpful discussions.  This work was supported in part by the U. S. Department of Energy under contract No.~DE-AC02-05CH11231 and No.~DE-AC02-98CH10886,
and the National Science Foundation under Grants No.~PHY-0969739 and No.~PHY-1316617.


\providecommand{\href}[2]{#2}\begingroup\raggedright\endgroup

\end{document}